\magnification=1100
\baselineskip=0.8truecm
\centerline{\bf Density Singularities and Cosmic Structures}
\medskip
\centerline{G. Murante, A. Provenzale, E. A. Spiegel$^{\dagger}$
and R. Thieberger$^*$}
\medskip
\centerline{Istituto di Cosmogeofisica, Corso Fiume 4, I-10133 Turin,
Italy}
\medskip
\centerline{$^\dagger$ normally at: Dept. of Astronomy, Columbia
University, New York, NY 10027, USA}
\medskip
\centerline{$^*$ normally at: Dept. of Physics, Ben Gurion
University, Beer Sheva, Israel}
\medskip
\medskip
\centerline{ \it{M.N.R.A.S., accepted} }
\noindent
{\bf Abstract}

\noindent
We show that an arrangement of singular density concentrations
accounts for the scaling exponents observed in the luminous
matter distribution in the Universe for scales below 10Mpc.
This model provides a good description of the matter distribution
on those smaller scales.

\medskip
\medskip
\noindent
{\bf 1. Introduction}

\noindent
Millennia ago, people looked at the stars in the sky and saw the
constellations.  Today we look at catalogues of galaxy positions
(Shane \& Virtanen 1967, Zwicky {\it et al.} 1968, Maddox {\it et
al.} 1990) and redshifts (Huchra {\it et al.} 1990, Haynes \& Giovanelli
1988, da Costa {\it et al.} 1989, Shectman {\it et al.} 1996) and
see groups, clusters and superclusters (de Vaucouleurs 1970, Bachall 1988)
and, on larger scales, voids, filaments and walls (see Jones 1992 for an
introductory review).  The big difference is in our present ability to
quantify the distribution and thus give meaning to the identification of
these structures.  Nevertheless, we still retain the desire to
rationalize what we see and there are several ways to go about this,
as described by Peebles (1980, 1992) and Borgani (1995) as well as authors
in Bonometto, Primack and Provenzale (1997).  The same quantitative
approaches may be applied to the results of
simulations of gravitationally interacting point masses in an
expanding universe, since these reveal a similar variety of structures.
Agreement between the analysis of observations and the computer results
encourages the belief that there may be an underlying, simple theoretical
picture of the evolution of the galaxy distribution.

We may also look for a rationalizing statistical picture and there
are two that now seem to be the main contenders for acceptance.
The first is the
conventional fractal picture (Mandelbrot 1982, Martinez 1991,
Provenzale 1991, Coleman \& Pietronero 1992, Borgani 1995) whose
antecedents are traceable to the past century.  This fits in well
with the belief that cosmic structures are formed in a cascade process
like that of turbulence theory (Thieberger, Spiegel \& Smith 1990,
Provenzale 1991, Provenzale {\it et al.} 1992) with a resulting rich
geometrical texture.  The second, equally compelling view, is that local
density singularities form and organize the matter distribution around
them (Peebles 1974, McClelland \& Silk 1977, Gurevich and Zybin 1988;
Sheth \& Jain 1996, White 1996).

Our aim in this paper is to scrutinize the singularity picture by testing
its ability to meet the quantitative tests and produce scalings like those
seen in the galaxy distribution.   That is, we shall find the generalized
dimensions of a set of singularities, just as one does in analyzing any
fractals.  As we shall see, the singularity picture is capable of
describing the scaling properties of the
galaxy distribution well on small scales of up to 10 Mpc.

\medskip\noindent
{\bf 2. Statistical descriptors.}

\noindent
Analyses of the spatial arrangement of galaxies have been based mainly
on both the correlation function (e.g., Peebles 1980) and
the correlation integral (Grassberger \& Procaccia 1983, Paladin \&
Vulpiani 1984).  The two-point correlation function, $\xi$, of a given
set of galaxies is defined such that the joint probability of finding a
galaxy in a volume element $\delta V_1$ and another in a volume element
$\delta V_2$ a distance $r$ away from the first is given by the expression
$\delta ^2P = n^2 \delta V_1\delta V_2 [1+\xi(r)]$, where $n$
is the mean number density of the sample.

For a uniform Poisson distribution, $\xi=0$ and the galaxies are
uncorrelated.  If $\xi$ is positive, the galaxies are said to be
correlated.  In the latter case, as $r$ goes to zero, $\xi$ becomes large
like an inverse power of $r$. This behaviour is expressed in the form $\xi
\propto r^{-\gamma}$ where $\gamma$ is a constant.  The precise value of
$\gamma$ depends on the morphological type of galaxies that are
considered and possibly on their luminosities. The value $\gamma
\approx 1.8$ has been found for $r$ less than five or ten $h^{-1}$
Mpc, especially from angular position catalogues (Peebles 1980),
where $h$ is the Hubble constant in units of 100 km/s/Mpc.

Higher order correlations may also be calculated, though the
formalism for such richer descriptions has been more actively
developed for the generalised correlation integrals.  In recent
elaborations of the theory of point sets
(see the reviews by Paladin \& Vulpiani 1987 and by Borgani 1995), one
introduces the moments
$$ C_q(r)= \left({1\over {N'}}\Sigma_i \left[
{1\over {N-1}} \Sigma_{j\ne i} \Theta(r - |{\bf X}_i - {\bf
X}_j|)\right]^{q-1}\right)^{1\over q-1} \eqno(1)$$
where $q$ is a parameter and
$\Theta$ is the Heaviside function. The inner summation is over the whole
set of $N-1$ galaxies with coordinates ${\bf X}_j$, $j\ne i$, and the
outer summation is over a subset of $N^{\prime}$ galaxies, taken as
centers, with coordinates ${\bf X}_i$.

To get an intuitive feel for this formula, consider ${\cal N}(r)$, the
average number of galaxies of the set lying within a distance $r$ of a
typical galaxy.  We see from the formula that $C_2(r)={\cal N}(r)/N$.
We may approximate ${\cal N}(r)$ as the integral,
$4\pi \int_0^r n(s) s^2 ds$,
where $n(r)$ is the density function.  The existence of structure in the
distribution of galaxies is connected to the dependence of $n$ on $r$ and
it may be shown that for a suitable constant $n_h$, we may write
$n(r)/n_h = 1+\xi(r)$ where $\xi(r)$ is the pair correlation function just
defined (Provenzale 1991). In the last few years, many lively
discussions have focussed on the question whether $n_h$ becomes independent
of the sample size above a certain threshold (see e.g. Lemson \&
Sanders 1991,  Coleman \& Pietronero 1992, 
Provenzale, Guzzo \& Murante 1994).

More generally, for any integer value of $q$, $C_q(r)$ is the fraction 
of $q$-tuples in the set whose members lie within a distance $r$ of one
another.  For sufficienly small $r$, $C_q$ will go to zero and, for a
typically well-behaved set, it will vanish like $r^{D_q}$, where the index
$D_q$ is called a generalized or Renyi dimension (Renyi 1970, Halsey {\it
et al.} 1986).  The quantity $C_2$ is called the correlation integral,
since it is the integral of $1+\xi$.  Consequently, by looking at
small $r$, we learn that $D_2=3-\gamma$ for those scales where $1+\xi(r)$
may be approximated by $\xi(r)$.

Since whatever scaling exists cannot be expected to be perfect, the values
obtained for $D_q$ (and $\gamma$) depend on the range of scales of $r$
under consideration.  To simplify the discussion, one may at the outset
limit the range of $r$ to include only portions of the range with a single
slope.  However, when the existence of a scaling range is not guaranteed
{\it a priori}, the practice of using a local slope is often followed.  In
this method, one plots $\log C_q$ against $\log r$ and determines the
local slope $D_q(r)$ as a function of the scale $r$.  In any range of $r$
for which $D_q(r)$ is nearly constant there is said to be scaling
behaviour for that particular range.

A certain amount of ambiguity in interpreting the observed galaxy
distribution may arise because some investigators try fit a power law
to $\xi$ while others prefer fitting one to $C_2$, thus to $1+\xi$.
Evidently, for small enough $r$, this makes no difference, but
discrepant results have been reported because practical considerations
sometimes force the range of $r$ included in the analysis to be
too large (Calzetti \& Giavalisco 1991).  Such discrepancies are
especially noticable in denser sets with strong self-overlap (Thieberger
{\it et al.} 1990), even in theoretical sets whose properties are
well controlled.  For real data, failure to restrict to very
small values of $r$ may aggravate the disagreements.
Nevertheless the earliest results giving $\gamma\approx
1.8$ seem robust for small enough $r$.  It is a
challenge to the theory to explain this number while
rationalizing the existence of the different scaling regimes.

Other results of the analysis of the galaxy distribution (e.g.,
Martinez \& Jones 1990, Borgani 1995) and of the output from numerical
simulations (Valdarnini, Borgani \& Provenzale 1992, Colombi, Bouchet \&
Scheffer 1992, Yepes, Dominguez-Tenreiro \& Couchman 1992, Murante {\it et
al.} 1996), confirm the existence of approximate scaling behaviour
with $1 < D_q < 1.4$ for $q \ge 2$, at scales below 5 $h^{-1}$ Mpc.
At larger scales, the existence of another scaling regime has been
suggested (Guzzo {\it et al.} 1991, Yepes {\it et al.} 1992, Murante {\it
et al.} 1996) with $D_2 \approx 2$. At still larger scales, a growth of
the generalized dimensions with $r$ is usually observed (Peebles 1989).
This behaviour of the generalized dimensions at the largest values of
$r$, with $D_q(r)$ increasing with $r$, is in general considered to
reflect an approach to homogeneity at large scales.  This is in keeping
with the smoothness of the background radiation, and most workers expect
the dimension to approach three, as for a space-filling distribution.
In any case, at the largest scales, other factors like time dependence and
curvature will have to be included before serious conclusions are reached.
For the richer structures seen at smaller scales, these problems do not
intrude and that is the regime we study here.

\medskip
\noindent{\bf 3. The Singularity Picture.}

\noindent Numerical simulations of structure formation are carried
out in an expanding background typically beginning at a state
that corresponds to the linearly evolved density field produced
at the time of recombination.  The initial perturbations to the uniform
state are often made as large as the observed fluctuations in the
background radiation allow.  The subsequent evolution shows the formation
of mass concentrations of various kinds --- flat structures, filaments and
clusters --- and generically leads to the development of a number of
local strong condensations (Efstahiou {\it et al.} 1988, Zurek and Warren
1994).  These structures form in the post-recombination era from density
perturbations that have survived either viscous and radiative damping and
dispersion due to random velocities.

The largest of the evolving structures are relatively immune
to the effects of pressure after recombination.  Any deviations from
sphericity in them will be exaggerated (see the literature cited in
Peebles 1980) as these early formations take on the flattened form known
as Zeldovich (1970) pancakes.  These arrange themselves in a complex
skein of sheet-like structures according to many simulations of this
process.  At the same time, as long as the pressure remains
unimportant, the
collapse will continue to form secondary structures, particularly at the
intersections of pancakes. As many discussions of the limiting behaviour of
these structures suggest, the density will locally evolve
toward the formation of singularities (Efstahiou {\it
et al.} 1988; Gurevich and Zybin 1988; Navarro, Frenk \& White 1995;
Cole \& Lacey 1996; White 1996).

The creation of local clusters seen in the real and simulated galaxy
distributions has focussed attention on spherical collapse.
This leads to the formation of singular densities in finite times
with a density $n(R)\propto R^{-\alpha}$ where $R$ is the distance from
the singularity.
Singularities of this form, with $\alpha$ between
1.5 and 3, are self-similar solutions of the Vlasov-Poisson equations
(Henriksen \& Widrow 1995).   Calculations by Gurevich \& Zybin (1988) for
special, but plausible, pressure-free
initial conditions lead to $\alpha=24/13$.
The case $\alpha=2$ corresponds to an isothermal gas
sphere, with pressure.
Tight spherical structures do seem to be the outcome of many simulations
of structure formation (Peebles 1980, White 1996) though there is no clear
argument for the existence of such spherical, presumably nongeneric,
structures.

We mention these theoretical notions as motivational and do not
assess their validity here.  Nor do we intend to compare their importance
to those of the other processes that have been considered in discussions
of structure formation, such as cosmic strings (e.g., Kolb \& Turner
1990) and fractal cascades (Provenzale, 1991).  For while it is true that
the power-law behaviour of the correlation integrals and the fractional
values of the generalized dimensions may reasonably be attributed to a
fractal nature of the galaxy distribution (Efstathiou, Fall \& Hogan 1979,
Mandelbrot 1982, Coleman, Pietronero \& Sanders 1988, Provenzale 1991,
Martinez 1991, Borgani 1995), we are interested here in the singularity
picture as a fruitful way to think about the galaxy statistics.

\medskip
\noindent{\bf 4. The Basic Scalings.}

\noindent We now show that the scaling behaviour of a random
superposition of singularities of a given type provides a good
description of the higher order statistical moments found in the galaxy
distribution.  These results extend previous calculations of the shape of
the correlation function $\xi$ for an isolated singularity
(Peebles 1974) or for a distribution of density singularities
(McClelland \& Silk 1977, Sheth \& Jain 1996).

Consider first the generalized dimensions of the set of points
disposed around a {\it single} condensation with power-law profile.
For large, positive $q$, the correlation integrals are dominated
by the very dense regions of the distribution near to the singularity
itself.  In the formula for $C_q$, the sum over $j$ gives the
number of galaxies within a
distance $r$ of the $i$th galaxy.  This quantity will clearly peak
sharply for a galaxy at (or very near to) the singularity itself
where it becomes equal to the mass interior to a sphere of
radius $r$ around the singularity.  The second sum averages
this quantity over the galaxies and it is completely dominated by
galaxies at the singularity.  Hence, for small $r$,
with $r\approx R$, the expression for $C_q(r)/[r^3 n(r)]$ approaches
a constant and does so most rapidly for $q\to \infty$. (Here $R$ is the
distance from the singularity and $r$ is the separation of points in the
statistical moment).

This argument shows that a singular density distribution, with
$n(R) \propto R^{-\alpha}$ for small $R$, has a scaling
behaviour (at least for small $r$) with $D_{\infty} = 3-\alpha$,
in analogy with the more familiar formula $D_2 = 3-\gamma$.
For any finite $q$, the value of the dimension $D_q$ is
larger than $3-\alpha$, because of the `dressing' of the
dimension by the contribution of points away from the singularity.
Nevertheless, for sufficiently strong singularities
($\alpha > 1.5$), we find a good scaling behaviour
even for moderate values of $q$.
In particular, using the result $\gamma=2\alpha-3$ of Peebles (1974),
one has that at small $r$ the value of $D_2$ tends to $D_2=3-\gamma=
6-2\alpha$ on those scales where $\xi \approx 1+\xi$.
In the singularity picture,
the scaling exponents, $D_q$, are thus largely determined by the singularity
exponent $\alpha$.

More explicitly, let points be put down in a three-dimensional
space at random with a probability proportional to the density
distribution $n(R) = n_0(R/R_{max})^{-\alpha}$ with
$R \le R_{max}$ and with $n(R)=0$
for $R > R_{max}$. To have finite masses in
such structures, we adopt finite values of $R_{max}$.
In calculating the correlation integrals to find $D_q(r)$ for this finite
sample of points, we face the difficulty that the
formula for $C_q(r)$ is not well adapted to
a set of galaxies with members near the edge
of the set.  So, in practise,
we base the evaluations on only a subset of the points whose members
are well away from the edges of the sample.  We then treat the
remaining points as background objects to complete the evaluations.
We do this for a variety of such subsets and average over the results
from each.  Using this approach, we obtain the results
shown in Fig. 1 for the singular distributions discussed above,
for the cases $\alpha = 1.8$, $2$ and $2.4$.
Scaling is clearly evident for
the moments with $q \ge 2$. For $q=2$, at small $r$ the
value of $D_2$ tends to $D_2=
6-2\alpha$, as expected from the theory.
For higher values of $q$, the dimensions become smaller and tend to
$D_\infty=3-\alpha$ for very large $q$.

For an arrangement of several such singular density concentrations, the
situation is like that in Fig. 2, which shows a random, uncorrelated
distribution of singularities with $\alpha=2$. The number density of
singularities is here $n_s=10$  and, for simplicity, we work
in a volume of unit size.  The average distance between two singularities
is $d=n_s^{-1/3}$.  The number of points around each singularity is
$N_0=100,000$ and we have chosen $R_{max}=0.8d$, so that different
singularities only mildly overlap.  We state the dimensions for this
case here and return in the next discussion to the question of the
effects of varying the parameters.

In Fig. 3 we show the generalized dimensions
calculated for the distribution of points of Fig. 2.
Good scaling behaviour is found, similar to that
found in the analysis of the real galaxy distribution
(Martinez \& Jones 1990, Borgani 1995) as well as in N-body
simulations (Valdarnini {\it et al.} 1992, Murante
{\it et al.} 1996).  When the singularities are placed
more deliberately, for instance on the sheet-like structures
produced by large-scale pancake formation, the qualitative appearance
of the distribution becomes even more plausible.  But an extensive
study of this latter issue is deferred to future work.  Before
turning to the meaning of these results we next examine how they
are affected when the parameters are assigned different values.

\medskip\noindent
{\bf 5. Role of the parameters}

\noindent
The form of the singularities discussed in the previous section is
$n(R)=n_0(R/R_{max})^{-\alpha}$. For a distribution of such singularities
we have to specify the values of four parameters: (1) the number density
of singularities $n_s$, or, equivalently, the average distance between two
singularities, $d=n_s^{-1/3}$; (2) the exponent in the power-law,
$\alpha$; (3) the cutoff radius, $R_{max}$, around a single density
enhancement; and (4) the density scale, $n_0=n(R_{max})$.  We also
have $N_0 = 4\pi\int_0^{R_{max}}n(r)r^2dr$, the number of points around
each singularity, which, in terms of the other parameters, is
$N_0=4\pi n_0 R_{max}^{3-\alpha}/(3-\alpha)$.

Clearly, $R_{max}$ and $d$ are not independently meaningful; the relevant
parameter is $\sigma=R_{max}/d$, which measures
how much overlap the
singularities have with each other.
When the singularity centers are too closely packed together, the
overlap becomes strong and it becomes increasingly hard to detect
scaling behaviour unless the number of points in the sample
is large enough to allow resolution of the cores of the
individual density concentrations.  This problem is common in dense sets
(Thieberger {\it et al.} 1990) and needs to be borne in mind
as we examine the role of the number of points $N_0$ per
singularity.  In Figure 4 we show the results of the scaling analysis
for the same distribution discussed in the previous section
($n_s=10$, $\alpha=2$, $\sigma=0.8$), but with $N_0=25,000$ (panel a)
and $N_0=7,000$ (panel b).  For the different values of $N_0$, no
difference is observed at large scales, while at small scales the lack of
statistics destroys the scaling behaviour.  This shows that varying $N_0$
(or $n_0$) for fixed $\alpha$, $\sigma$ and $n_s$ does not influence the
scale of the transition to homogeneity but only the detectability of the
small-scale scaling regime.  If $N_0$ is too small, there is the danger
that the small-scale regime is completely lost owing to the lack of
statistics.  Analogous results have been discussed by Borgani {\it et al.}
(1993) in the case of fractal distributions.

Next we turn to the role of the power-law exponent, $\alpha$.
Figure 5 shows the results of the scaling analysis for a distribution
of singularities with $n_s=10$ and $\alpha$ uniformly distributed between
1.6 and 2.4. Here $N_0=100,000$ and $\sigma=0.8$. For low values of
$q$, the scaling is less good due to the averaging over contributions
from different singularities. For larger $q$'s, the correlation integrals
give more and more weight to the most overdense regions, that are
associated with the stronger singularities. At these values of $q$,
scaling is observed with a dimension close to that of the singularity
with the largest values of $\alpha$.

The measure of overlap between concentrations is $\sigma=R_{max}/d$ and,
for galaxies, this will be determined by how long the density
perturbations have been evolving.  If $\sigma$ is quite small, the
singularities are isolated from each other and wide empty spaces remain
between them.  In this case, the scaling analysis indicates a
lack of points at intermediate scales. On the other hand, if $\sigma$ is
too large many points are in the overlapping portions with a corresponding
lack of statistics at small scales. In this case, very large values of
$N_0$ are needed in order to detect the small-scale regime. A value
$\sigma=0.8$ provides very little overlap without leaving great empty
spaces.  However, the precise value of $\sigma$ is not crucial to the
scaling results.  Figure 6 shows such results for a distribution of
density singularities with $n_s=10$, $\alpha=2$, $N_0=100,000$ and
$R_{max}$ uniformly distributed between $0.5d$ and $1.5d$. The results
are very similar to those obtained with $R_{max}=0.8d$.

Finally, we note that some investigations of the results of cosmological
N-body simulations have indicated that large density peaks could produce
behaviour other than pure power laws.   In fact, if the time of
singularity has not been achieved, there will be density maxima that are
rounded in their inner cores.  One form of density maximum that has been
considered is (Navarro et al, 1995) $ n(R) = (R_0/R)(1+R/R_s)^{-\nu}$,
where $R_0$ is a normalization factor and $R_s$ marks the transition from
the $R^{-1}$ behaviour at small scales to the $R^{-\nu-1}$ behaviour at large
scales. Typically, $\nu=3$ for the Hernquist model and $\nu=2$ for the NFW
model (Navarro et al, 1995).  Our analysis shows that for an isolated
density peak with this shape, no clear scaling behaviour is observed.
Independently of whether we understand the origin of power law behaviour
in the density, it does seem to be in line with the observations of
large-scale structure.

\medskip\noindent
{\bf 6. Conclusions}
\noindent

The aim of this work was to see how well an arrangement 
of density singularities
can account for the observed scaling of the galaxy
distribution.   We conclude that it does this as well as the conventional
picture of a fractal distribution (in which we include the multifractal
as well).  Of course, one might argue that, since the dimensions we find
are not integers, the singularity picture is also a fractal one and,
on this matter, we would not disagree.  Nevertheless, we believe that it
is
fair to claim that an explanation of the observed scaling in terms of
singular densities is qualitatively different from a conventional
geometric fractal picture.

On physical grounds, the singularity scenario is well justified
and it is supported by N-body simulations that indicate the existence
of strong density enhancements at small scales.  On this view,
the origin of the scaling exponents for small scales in the present data
is to be found in the structure of the dominant singularity with the value
$\gamma=1.8$ corresponding to $\alpha=2.4$ (Peebles, 1974).
At larger scales, the spatial distribution of singularities
themselves becomes important (Sheth \& Jain 1996).

The existence of scaling behaviour in the cluster distribution
(Bachall 1988, Borgani 1995) may indicate a geometrical arrangement of
the small scale condensations.  This could point to a truly fractal
distribution of galaxy clusters or higher structures on large scales.  On
the other hand, this regime may result from another type of singularity,
such as pancakes. Work is now in progress to properly address this issue
in the framework of the singularity picture and to study the
dependence of the density profile of the singularities on the
initial conditions.

But it is not the scaling distribution alone that decides which
description to invoke in trying to learn whether large scale structure
is formed mainly by cascades or by collapse onto singularities.
There are other observational
features indicating that singularities are prevalent on the smaller
scales.  Recent detection of cuspy density distributions that seem
describable as isothermal spheres suggest singular density distributions
with $\alpha \approx 2$ (Fukushige \& Makino 1996).  Also nearly flat
galactic rotation curves seem compatible with the present estimates of
$\alpha$.

For a variety of disk galaxies, the rotation curves vary like
$r^{\epsilon}$ with $\epsilon$ of the order of $\pm 0.1$ (Begeman {\it et
al.} 1991).  On the other
hand, for the spherical singular density distribution we are
considering, the rotational velocity will vary like $r^{(2-\alpha)/2}$.
This gives us values of $\epsilon=(2-\alpha)/2$
in the observational range for $\alpha = 2\pm 0.2$.
The agreement in order of magnitude is suggestive of a universality
of density singularities.  There is much to be understood in the
galaxy data, including the correlation of $\epsilon$ with other
properties of the galaxies, but this coincidence of values seems worth
pondering.

We conclude that the presence of density singularities is likely to be a
common feature of the large scale structure of the universe and that the
study of the evolution of such singularities will continue to be a fruitful
activity.
\bigskip
\noindent We thank Dr. R.K. Sheth for his through reading of the
manuscript. We acknowledge B. Dubrulle, J. v. Gorkom, I. Grenier and A.
Refregier for pointing out relevant references.

\vfill
\eject
\noindent{ \bf References.}

\noindent Bachall, N., 1988, Ann. Rev. Astron. Astrophys., 26, 631.

\noindent Begeman, K.G., Broeils, A.H. \& Sanders, R.H., 1991, MNRAS,
249, 523.

\noindent Bonometto, S., Primack, J. \& Provenzale, A., Eds., 1997,
{\it Dark Matter in the Universe}, Proceedings of the CXXXII Course of
the International School of Physics ``E. Fermi" (IOP Press).

\noindent Borgani, S., Murante, G., Provenzale, A. \& Valdarnini, R.,
1993, Phys. Rev. E, 47, 3879.

\noindent Borgani, S., 1995, Phys. Rep., 251, 1.

\noindent Calzetti, D. \& Giavalisco, M., 1991, in {\it Applying Fractals
in Astronomy}, A. Heck and J.M. Perdang Eds. (Berlin: Springer).

\noindent Cole, S. M. \& Lacey, C. G., 1996, MNRAS, in press;
preprint astro-ph/9510047.

\noindent Coleman, P.H., Pietronero, L. \& Sanders, R.H., 1988, A\&A,
200, L32.

\noindent Coleman, P.H. \& Pietronero, L., 1992, Phys. Rep.,
231, 311.

\noindent Colombi, S., Bouchet, F.R. \& Scheffer, R., 1992, A\&A,
263, 1.

\noindent da Costa, L.N. {\it et al.}, 1989, AJ, 97,
315.

\noindent de Vaucouleurs, G., 1970, Science, 167, 1203.

\noindent Efstathiou, G., Fall, S.M. \& Hogan, C., 1979, MNRAS, 189,
203.

\noindent Efstathiou, G., Frenk, C.S., White, S.D.M. \&
Davis, M., 1988, MNRAS, 235, 715.

\noindent Fukushige, T. \& Makino, J., 1996, preprint astro-ph/9610005.

\noindent Gurevich, A. V. \& Zybin, P., 1988, Sov. Phys. JETP, 67, 1957.

\noindent Guzzo, L., Iovino, A., Chincarini, G., Giovanelli, R. \&
Haynes, M.P., 1991, ApJLett, 382, L5.

\noindent Grassberger, P. \& Procaccia, I., 1983, Phys. Rev. Lett.,
50, 346.

\noindent Halsey, T.C., Jensen, M.H., Kadanoff, L.P., Procaccia,
I. \& Shraiman, B.I., 1986, Phys. Rev. A, 33, 1141.

\noindent Haynes, M.P. \& Giovanelli, R., 1988, in {Large Scale Motions
in the Universe}, V. Rubin \& G.V. Coyne Eds.
(Princeton: Princeton Univ. Press).

\noindent Huchra, J.P., Geller, M. J., de Lapparent, V. \& Corwin,
H.G.Jr, 1990, ApJ Suppl., 72, 433.

\noindent Jones, B.J.T., 1992, in {\it Observational and Physical
Cosmology}, F. Sanchez, M. Collados \& R. Rebolo Eds.
 (Cambridge: Cambridge Univ. Press).

\noindent Kolb, E.W. \& Turner, M.S., 1990, {\it The Early Universe}
(California: Addison-Wesley).

\noindent Lemson, G., \& Sanders,
R.H., 1991, MNRAS, 252, 319.

\noindent Maddox, S.J., Efstathiou, G., Sutherland, V.J. \& Loveday,
J., 1990, MNRAS, 242, 43p.

\noindent Mandelbrot B.B., 1982, {\it The Fractal Geometry of Nature}
(San Francisco: Freeman).

\noindent Martinez, V.J. \& Jones, B.J.T., 1990, MNRAS, 242, 517.

\noindent Martinez, V.J., 1991, in
{\it Applying Fractals in Astronomy},
A. Heck and J.M. Perdang Eds. (Berlin: Springer).

\noindent McClelland, J. \& Silk, J., 1977, ApJ, 217, 331.

\noindent Murante, G., Provenzale, A., Borgani, S., Campos, A.
\& Yepes, G., 1996, {\it Astroparticle Phys.}, 5, 53.

\noindent Navarro, J.F., Frenk, C.S. \& White, S.D.M., 1995, MNRAS,
275, 720.

\noindent Paladin G. \& Vulpiani A., 1984, Nuovo Cimento Lett., 41,
82.

\noindent Paladin G. \& Vulpiani A., 1987, Phys. Rep., 156, 147.

\noindent Peebles, P.J.E., 1974, A\&A, 32, 197.

\noindent Peebles P.J.E., 1980, {\it The Large Scale Structure of the
Universe} (Princeton: Princeton Univ. Press).

\noindent Peebles, P.J.E., 1989, Physica D, 38, 273.

\noindent Peebles, P.J.E., 1992, {\it Principles of Physical Cosmology}
(Princeton: Princeton Univ. Press).

\noindent Provenzale, A., 1991, in {\it Applying Fractals in Astronomy},
A. Heck and J.M. Perdang Eds. (Berlin: Springer).

\noindent Provenzale, A., Galeotti, P., Murante, G. \& Villone, B., 1992,
ApJ, 401, 455.

\noindent Provenzale, A., Guzzo, L. \& Murante, G., 1994,
MNRAS, 266, 55.

\noindent Renyi, A., 1970, {\it Probability Theory} (Amsterdam: North-Holland).

\noindent Shane, C.D. \& Virtanen, C.A., 1967, Publ. Lick Obs., 22.

\noindent Shectman, S.A., Landy, S.D., Oemler, A., Tucker, D.L.,
Lin, H., Kirshner, R.P, Schechter, P.L., 1996, ApJ, in press;
preprint astro-ph/9604167.

\noindent Sheth, R.K. \& Jain, B., 1996, submitted to MNRAS;
preprint astro-ph/9602103.

\noindent Thieberger, R., Spiegel, E.A. \& Smith, L.A., 1990, in
{\it The Ubiquity of Chaos}, S. Krasner Ed. (AAAS Press).

\noindent Valdarnini, R., Borgani, S. \& Provenzale, A., 1992,
ApJ, 394, 422.

\noindent Yepes, G., Dominguez-Tenreiro, R. \& Couchman, H.M.P., 1992,
ApJ, 401, 40.

\noindent White, S.D.M., 1996, in {\it Gravitational Dynamics}, O. Lahav,
E. Terlevich \& R. Terlevich Eds. (Cambridge: Cambridge Univ. Press).

\noindent Zel'dovich, Ya.B., 1970, A\&A, 4, 84.

\noindent Zurek, W.H. \& Warren, M.S., 1994, Los Alamos Science, No. 22,
58.

\noindent Zwicky, F., Herzog, E., Karpowicz, M. \& Koval, C.T., 1968,
{\it Catalogue of Galaxies and Clusters of Galaxies}
(Pasadena: California Inst. of
Technology).

\vfill
\eject
\noindent {\bf Figure Captions.}
\medskip
\noindent Figure 1.
Effective generalized dimensions for $q=2,4,6$ for points placed
at random accordingly to the density $n(R) = R^{-\alpha}$,
for $\alpha=1.8$ (panel a), $\alpha=2$ (panel b) and
$\alpha=2.4$ (panel c).
Dimensions are obtained by averaging
over ten resamplings of the subsets
used as centers; error
bars are the $3\sigma$ deviations over the resamplings.
\medskip
\noindent Figure 2.
Point distribution sampling the density field generated by
a random distribution of
10 singularities in the unit cube with size $L=1$. Each singularity
has a shape $n(R) = n_0 (R/R_{max})^{-\alpha}$, here
$\alpha=2$  and $R_{max}=0.8d$ where $d=10^{-1/3}$ is the average distance
between two singularities. The number of points per singularity is
$N_0=100,000$.
\medskip
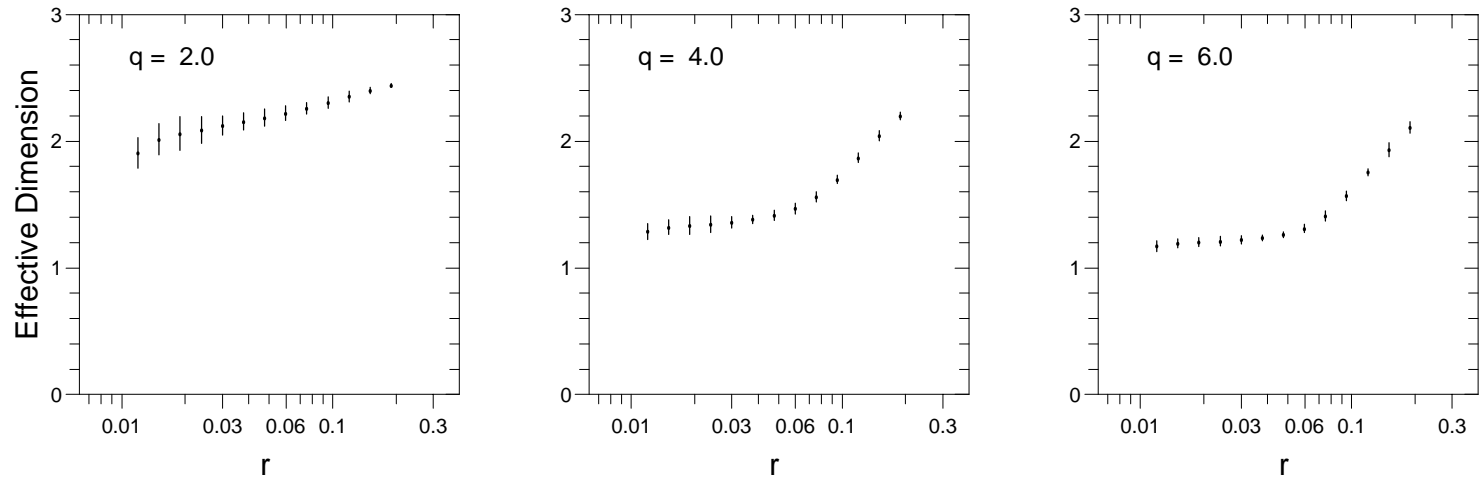
\noindent Figure 3.
Effective generalized dimensions for $q=2,4,6$ for the
point distribution shown in Figure 2.
\medskip
\noindent Figure 4.
Effective generalized dimensions for $q=2,4,6$ for a
distribution similar to that shown in Figure 2, but with
$N_0=25,000$ (panel a) and $N_0=7,000$ (panel b).
\medskip
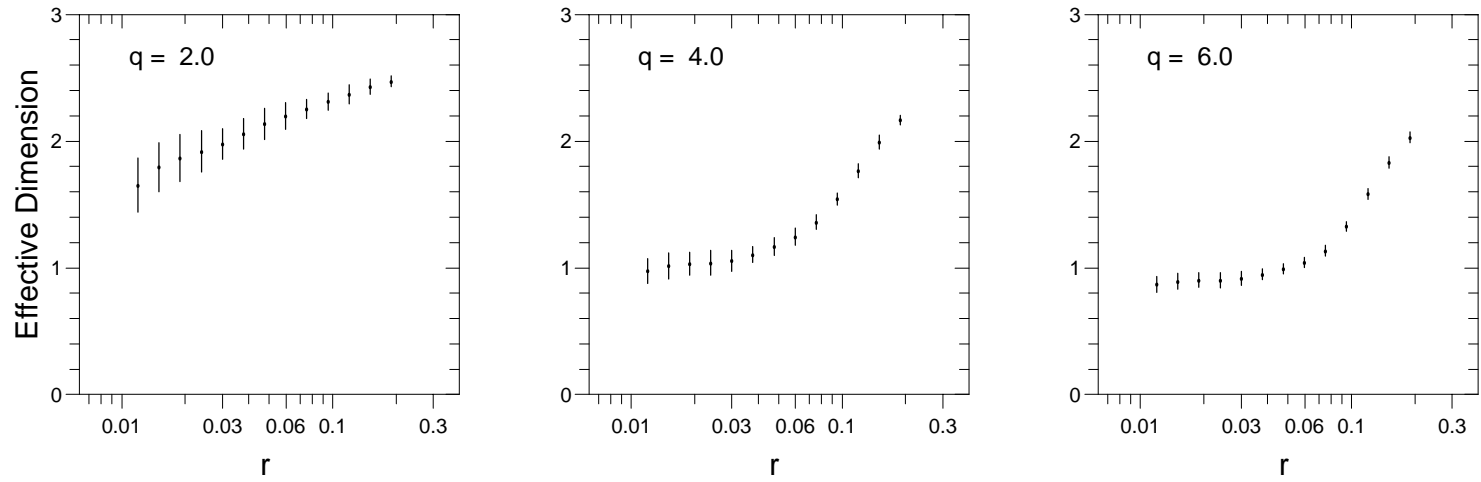
\noindent Figure 5.
Effective generalized dimensions for $q=2,4,6$ for a
distribution of 10 singularities with $\alpha$ uniformly
distributed between 1.6 and 2.4. Here $R_{max}=0.8d$ and
$N_0=100,000$.
\medskip
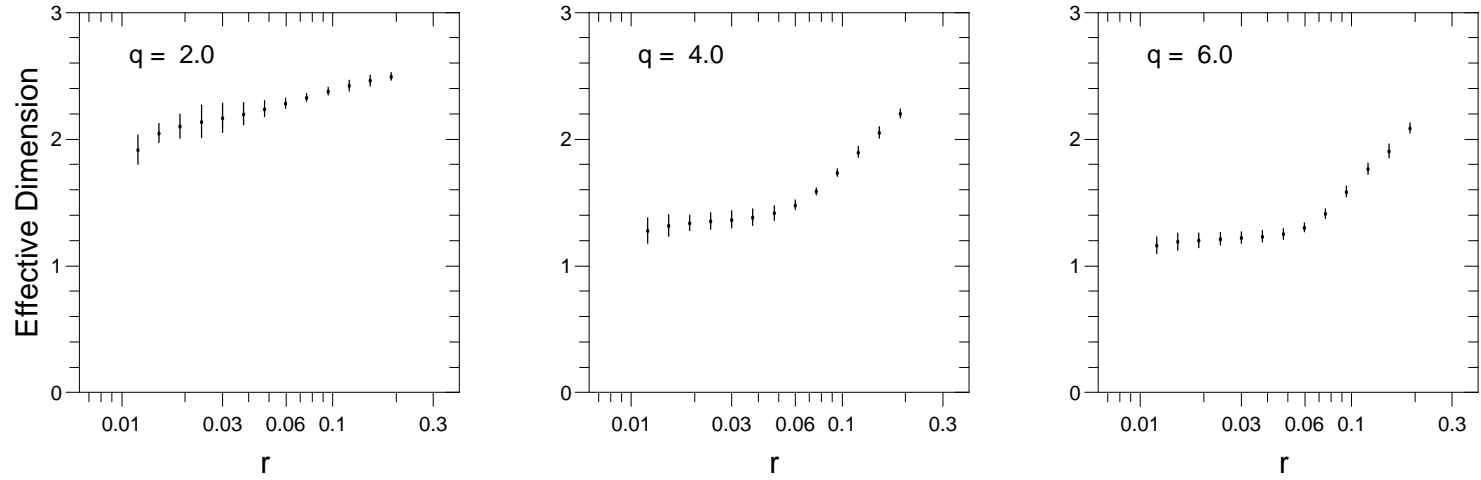
\noindent Figure 6.
Effective generalized dimensions for $q=2,4,6$ for a
distribution of 10 singularities with $\alpha=2$ and
$R_{max}$ uniformly distributed between $0.5d$ and $1.5d$.
Here $N_0=100,000$.
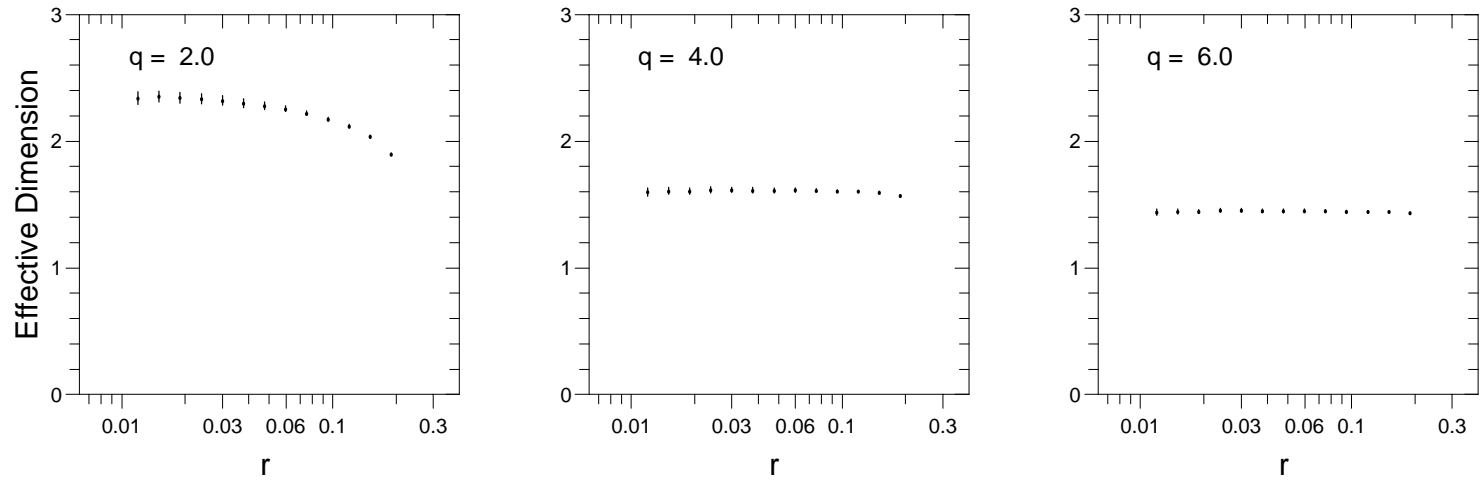
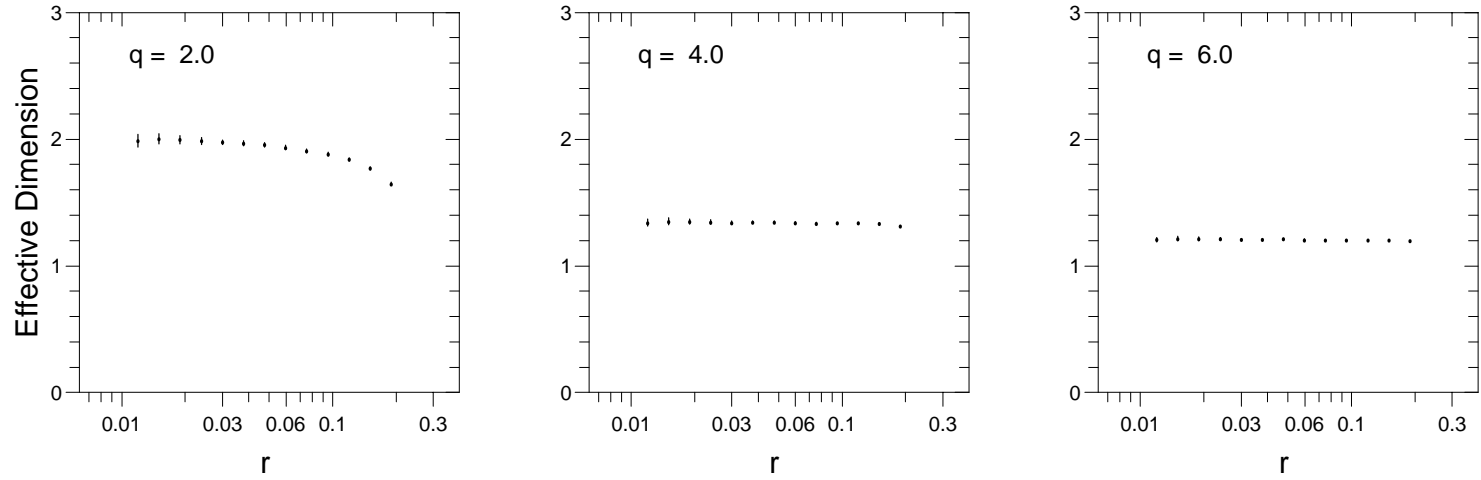
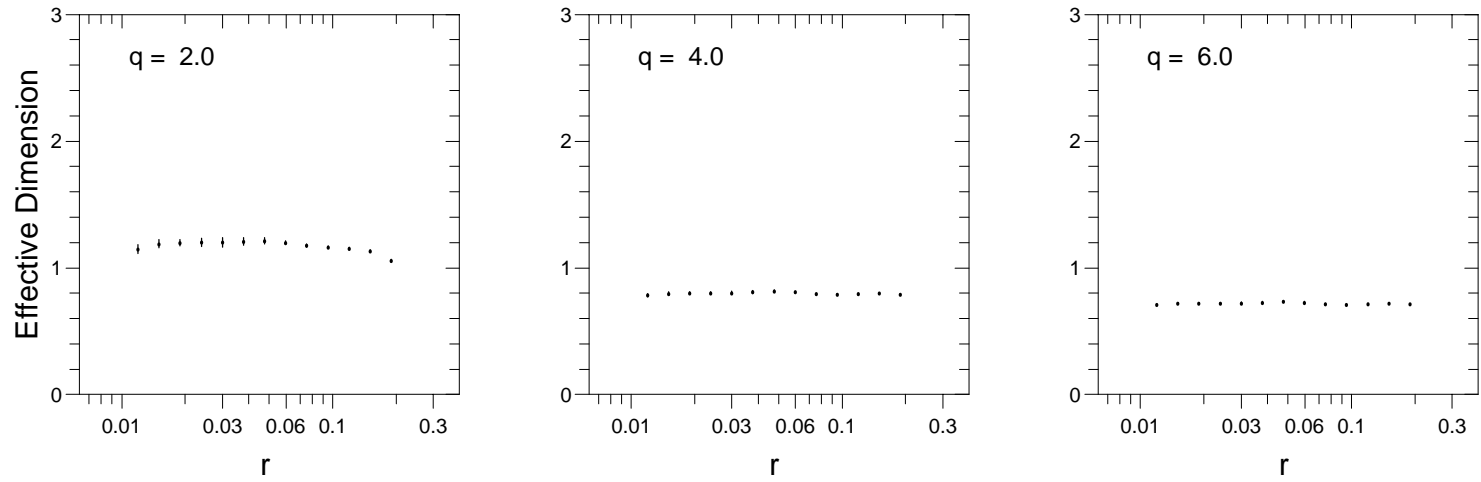
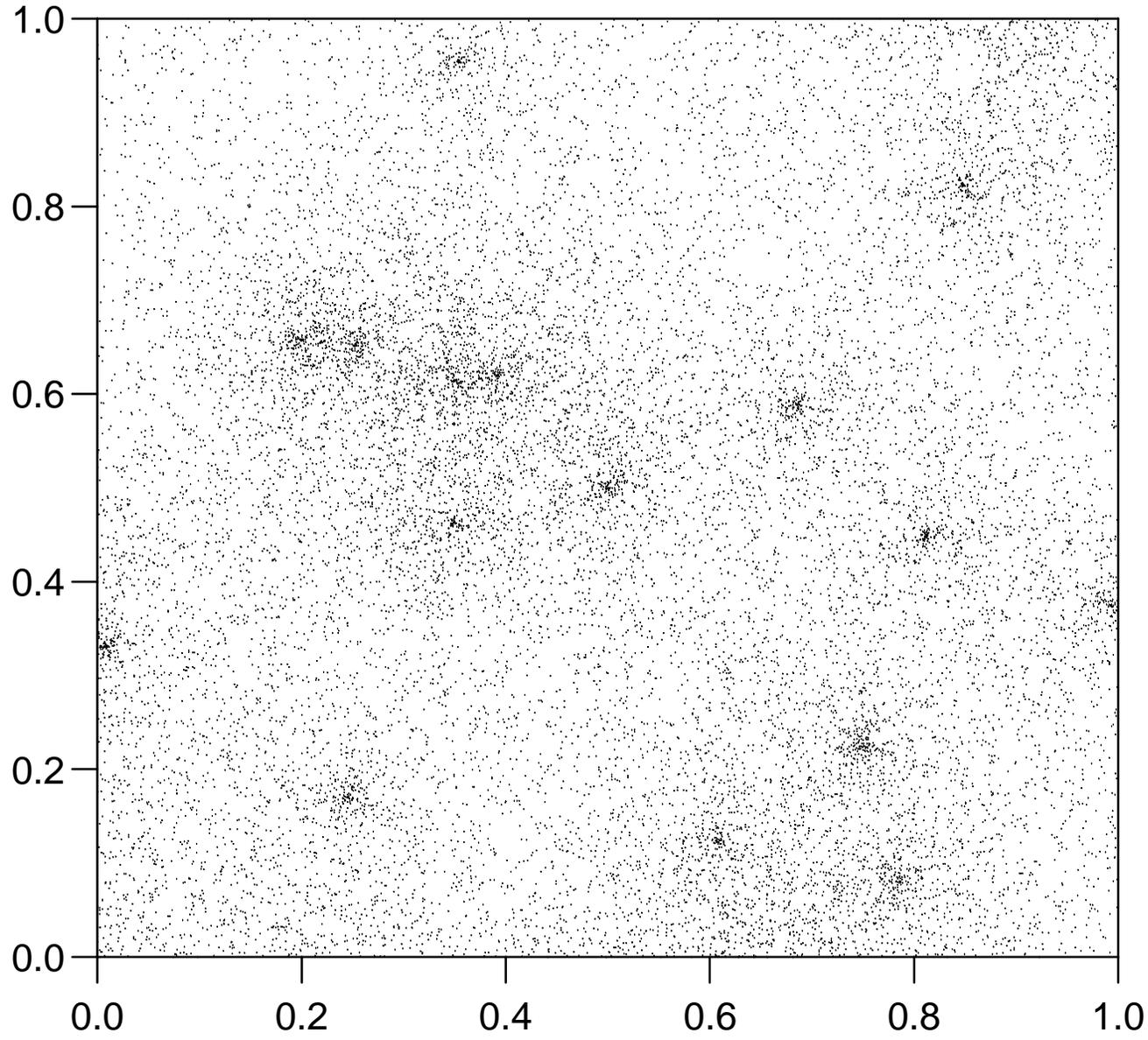
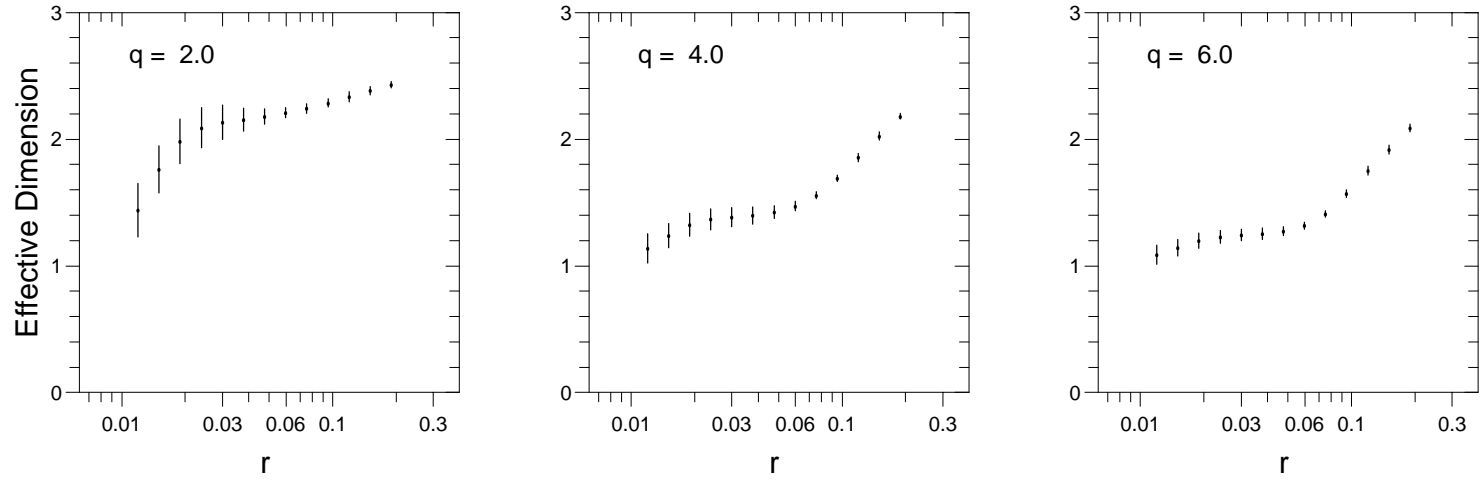
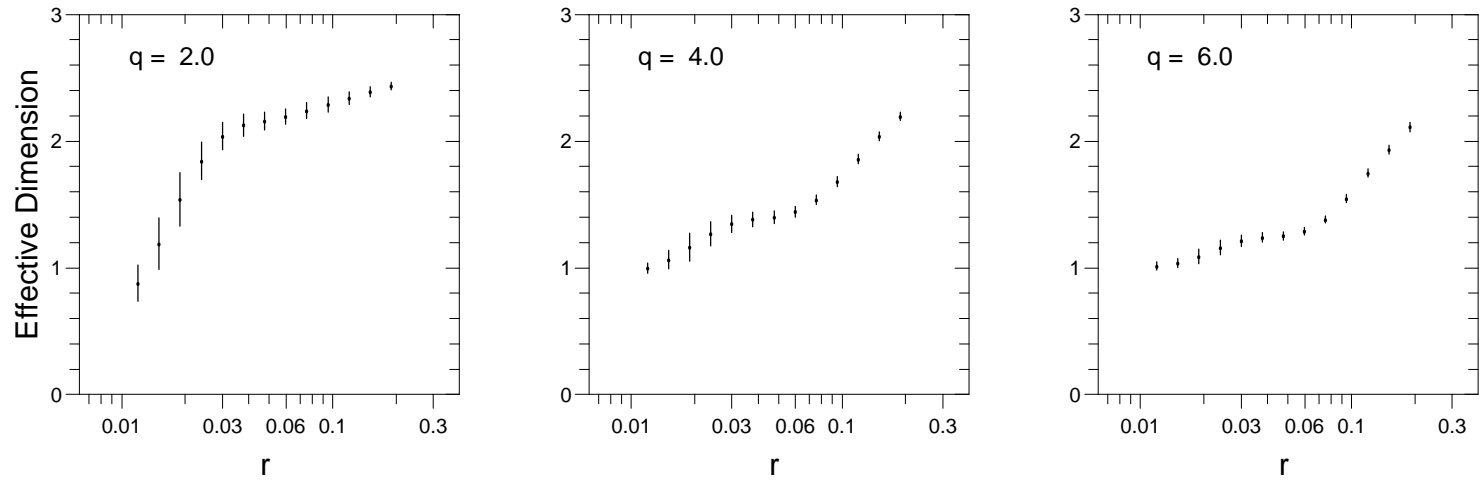
\end